\documentclass[sigconf]{acmart}

\usepackage{hyphenat}
\usepackage{xspace}
\usepackage{amsmath}
\usepackage{amsfonts}
\usepackage{url}
\usepackage{booktabs}
\usepackage{multirow}
\usepackage{makecell}
\usepackage{caption}
\usepackage{minibox}
\usepackage{graphicx}
\usepackage{mathtools}
\usepackage{color}
\usepackage{ifthen}
\usepackage{textcomp}
\usepackage{enumitem}
\usepackage{verbatim}
\usepackage{algorithm}
\usepackage{algorithmic}

\usepackage{amsthm}
\theoremstyle{plain}

\newcommand{\chatoDisplayMode}[1]{#1}

\definecolor{MyRed}{rgb}{0.6,0.0,0.0} 
\definecolor{MyBlack}{rgb}{0.1,0.1,0.1} 
\definecolor{MyGray}{rgb}{0.5,0.5,0.5} 
\newcommand{\inred}[1]{{\color{MyRed}\sf\textbf{\textsc{#1}}}}
\newcommand{\frameit}[2]{
  \begin{center}
  {\color{MyRed}
  \framebox[.9\columnwidth][l]{
    \begin{minipage}{.85\columnwidth}
    \inred{#1}: {\sf\color{MyBlack}#2}
    \end{minipage}
  }\\
  }
  \end{center}
}

\newcommand{\note}[2][]{\chatoDisplayMode{\def\@tmpsig{#1}\frameit{{\Pointinghand} Note}{#2\ifx \@tmpsig \@empty \else \mbox{ --\em #1}\fi}}}
\newcommand{\todo}[2][]{\chatoDisplayMode{\def\@tmpsig{#1}\frameit{{\Writinghand} To-do}{#2\ifx \@tmpsig \@empty \else \mbox{ --\em #1}\fi}}}

\newcommand{\abbrevStyle}[1]{#1}

\newcommand{\eg}{\abbrevStyle{e.g.}\xspace}

\newcommand{\etal}{\abbrevStyle{et al.}\xspace}
\newcommand{\vs}{\abbrevStyle{vs.}\xspace}
\newcommand{\etc}{\abbrevStyle{etc.}\xspace}

\newcommand{\Secref}[1]{Sec.~\ref{#1}}

\newcommand{\Figref}[1]{Fig.~\ref{#1}}

\newcommand{\xhdr}[1]{\vspace{1.7mm}\noindent{{\bf #1.}}}

\newcommand{\textcite}[1]{\citeauthor{#1} \shortcite{#1}}

\newcommand{\hide}[1]{}

\makeatletter
\newcommand{\iffont}[2]{\ifthenelse{\equal{\f@family}{#1}}{#2}{}}
\makeatother

\iffont{ptm}{
  \usepackage{mathptmx}

  \DeclareSymbolFont{greek}{OML}{cmm}{m}{n}
  \DeclareMathSymbol{\alpha}{\mathalpha}{greek}{"0B}
  \DeclareMathSymbol{\beta}{\mathalpha}{greek}{"0C}
  \DeclareMathSymbol{\gamma}{\mathalpha}{greek}{"0D}
  \DeclareMathSymbol{\delta}{\mathalpha}{greek}{"0E}
  \DeclareMathSymbol{\epsilon}{\mathalpha}{greek}{"0F}
  \DeclareMathSymbol{\zeta}{\mathalpha}{greek}{"10}
  \DeclareMathSymbol{\eta}{\mathalpha}{greek}{"11}
  \DeclareMathSymbol{\theta}{\mathalpha}{greek}{"12}
  \DeclareMathSymbol{\iota}{\mathalpha}{greek}{"13}
  \DeclareMathSymbol{\kappa}{\mathalpha}{greek}{"14}
  \DeclareMathSymbol{\lambda}{\mathalpha}{greek}{"15}
  \DeclareMathSymbol{\mu}{\mathalpha}{greek}{"16}
  \DeclareMathSymbol{\nu}{\mathalpha}{greek}{"17}
  \DeclareMathSymbol{\xi}{\mathalpha}{greek}{"18}
  \DeclareMathSymbol{\pi}{\mathalpha}{greek}{"19}
  \DeclareMathSymbol{\rho}{\mathalpha}{greek}{"1A}
  \DeclareMathSymbol{\sigma}{\mathalpha}{greek}{"1B}
  \DeclareMathSymbol{\tau}{\mathalpha}{greek}{"1C}
  \DeclareMathSymbol{\upsilon}{\mathalpha}{greek}{"1D}
  \DeclareMathSymbol{\phi}{\mathalpha}{greek}{"1E}
  \DeclareMathSymbol{\chi}{\mathalpha}{greek}{"1F}
  \DeclareMathSymbol{\psi}{\mathalpha}{greek}{"20}
  \DeclareMathSymbol{\omega}{\mathalpha}{greek}{"21}
  \DeclareMathSymbol{\varepsilon}{\mathalpha}{greek}{"22}
  \DeclareMathSymbol{\vartheta}{\mathalpha}{greek}{"23}
  \DeclareMathSymbol{\varpi}{\mathalpha}{greek}{"24}
  \DeclareMathSymbol{\varrho}{\mathalpha}{greek}{"25}
  \DeclareMathSymbol{\varsigma}{\mathalpha}{greek}{"26}
  \DeclareMathSymbol{\varphi}{\mathalpha}{greek}{"27}
  \DeclareSymbolFont{otone}{OT1}{cmr}{m}{n}
  \DeclareMathSymbol{\Gamma}{\mathalpha}{otone}{0}
  \DeclareMathSymbol{\Delta}{\mathalpha}{otone}{1}
  \DeclareMathSymbol{\Theta}{\mathalpha}{otone}{2}
  \DeclareMathSymbol{\Lambda}{\mathalpha}{otone}{3}
  \DeclareMathSymbol{\Xi}{\mathalpha}{otone}{4}
  \DeclareMathSymbol{\Pi}{\mathalpha}{otone}{5}
  \DeclareMathSymbol{\Sigma}{\mathalpha}{otone}{6}
  \DeclareMathSymbol{\Upsilon}{\mathalpha}{otone}{7}
  \DeclareMathSymbol{\Phi}{\mathalpha}{otone}{8}
  \DeclareMathSymbol{\Psi}{\mathalpha}{otone}{9}
  \DeclareMathSymbol{\Omega}{\mathalpha}{otone}{10}
  \DeclareSymbolFont{syms}{OML}{cmm}{m}{it}
  \DeclareMathSymbol{\partial}{\mathord}{syms}{"40}
  \DeclareMathAlphabet{\mathbold}{OML}{cmm}{b}{it}
  \DeclareSymbolFont{largesymbols}{OMX}{cmex}{m}{n}

}

\AtBeginDocument{%
  \providecommand\BibTeX{{%
    \normalfont B\kern-0.5em{\scshape i\kern-0.25em b}\kern-0.8em\TeX}}}

\setcopyright{none}
\renewcommand\footnotetextcopyrightpermission[1]{} 
\acmConference{Pre-print}
\acmBooktitle{}
\acmPrice{}
\acmISBN{}

\begin{document}

\title{Are Anti-Feminist Communities Gateways to the Far Right?
Evidence from Reddit and YouTube}

\author{Robin Mamié}
\affiliation{
\institution{EPFL}
\country{Switzerland}
}
\email{robin.mamie@epfl.ch}

\author{Manoel Horta Ribeiro}
\affiliation{
\institution{EPFL}
\country{Switzerland}
}
\email{manoel.hortaribeiro@epfl.ch}

\author{Robert West}
\affiliation{
\institution{EPFL}
\country{Switzerland}
}
\email{robert.west@epfl.ch}

\renewcommand{\shortauthors}{Mamié et al.}

\begin{abstract}
Researchers have suggested that ``the Manosphere,'' a conglomerate of men-centered online communities, may serve as a gateway to far right movements.
In that context, this paper quantitatively studies the migratory patterns between a variety of groups within the Manosphere and the Alt-right, a loosely connected far right movement that has been particularly active in mainstream social networks. 
Our analysis leverages over 300 million comments spread through Reddit (in 115 subreddits) and YouTube (in 526 channels) to investigate whether the audiences of channels and subreddits associated with these communities have converged between 2006 and 2018.
In addition to subreddits related to the communities of interest, we also collect data on counterparts: other groups of users which we use for comparison (e.g., for YouTube we use a set of media channels).
Besides measuring the similarity in the commenting user bases of these communities, we perform a migration study, calculating to which extent users in the Manosphere gradually engage with Alt-right content.
Our results suggest that there is a large overlap between the user bases of the Alt-right and of the Manosphere and that members of the Manosphere have a bigger chance to engage with far right content than carefully chosen counterparts.
However, our analysis also shows that migration and user base overlap varies substantially across different platforms and within the Manosphere.
Members of some communities (e.g., Men's Rights Activists) gradually engage with the Alt-right significantly more than counterparts on both Reddit and YouTube, whereas for other communities, this engagement happens mostly on Reddit (e.g., Pick Up Artists).
Overall, our work paints a nuanced picture of the pipeline between the Manosphere and the Alt-right, which may inform platforms' policies and moderation decisions regarding these communities. 
\end{abstract}




\settopmatter{printacmref=false}

\maketitle

$*$ This paper has been accepted at the 13th ACM Web Science Conference (WebSci'21), please cite accordingly.

\newpage

\section{Introduction}

A conglomerate of online communities broadly referred to as ``the Manosphere'' has received increased attention from scholars (\eg, \cite{ribeiro_evolution_2020, ging_alphas_2017}) and the media (\eg, \cite{emba_men_2019, sharlet_are_2014}).
According to Lily~\cite{lilly_world_2016}, these communities are united by a belief in a ``crisis in masculinity'' caused by feminists and feminist ideology.
Although the roots of these communities can be traced back to the Men's Rights Movement created in the 1970s~\cite{messner_limits_1998},
recent instances of online harassment~\cite{jones_sluts_2020} and real-world violence~\cite{dewey_inside_2014} highlight their impact on society.

As these communities flourished on mainstream and fringe platforms~\cite{ribeiro_evolution_2020}, a chief concern that emerged from researchers is the link between the Manosphere and other fringe groups, including White Supremacist and Identitarian movements~\cite{jackson_schema_2019}.
These connections would make the Manosphere fertile recruiting grounds for known violent extremist groups, but also suggest the emergence of a new form of extremism~\cite{dibranco_male_2020}.
Not coincidentally, in 2018, the term ``Male Supremacy'' started being tracked as an ideology by both the Southern Poverty Law Center (SLPC)~\cite{slpc_male_2018} and the National Consortium for the Study of Terrorism and Responses to Terrorism (START)~\cite{start_profiles_2018}.

The notion that the heterogeneous roster of Manosphere-related communities would act as a ``gateway'' or a ``pipeline'' to fringe world views resonates with previous research on the role of contrarian communities on YouTube.
According to Lewis~\cite{lewis_alternative_2018}, the proximity between content creators on YouTube---which not only share the same platform but often engage in public debate---could create ``radicalization pathways'' on the website.
These radicalization pathways were quantitatively studied by Ribeiro \etal~\cite{ribeiro_auditing_2020}, who showed that members of other online communities, such as the so-called Alt-lite, consistently migrated to channels espousing more extreme views.

In this context, the importance of quantitatively studying the migration patterns between the Manosphere and the far right is twofold.
First, it contributes to the understanding of fringe communities and their connections.
Previous work has mapped the evolution of the different groups within the Manosphere~\cite{ribeiro_evolution_2020}, producing a data-driven history of these (mostly anonymous) communities.
In a similar fashion, understanding alleged migrations from Manosphere communities to far right groups can help us make sense of our complicated online ecosystem.
Second, it is a continuation of an effort to identify and characterize communities that are gateways to extremism~\cite{lewis_alternative_2018, ribeiro_auditing_2020}.
Although qualitative evidence for the link between the Manosphere and the far right are plenty~\cite{jackson_schema_2019, dibranco_male_2020},
we argue that it is important to quantify the strength of this link and to compare it with previous known gateway communities (\eg, the Alt-lite).
The Manosphere is a wide and heterogeneous set of communities, which range from Men's Rights Activists (MRAs) who claim that society is rigged against men to Involuntary Celibates (Incels) who believe to be destined to a life of romantic rejection.
Different groups within the Manosphere may have different links to the far right, and quantifying these ties may precisely inform platforms' policies and moderation decisions. 

\noindent
\textbf{Present work}
In this paper, we investigate the link between various communities in the Manosphere\footnote{We use Lilly's~\cite{lilly_world_2016} division of the Manosphere which divides the manosphere into four broad groups: Men Going Their Own Way (MGTOWs), Men's Rights Activists (MRAs), Involuntary Celibates (Incels) and Pick Up Artists (PUAs).} and the Alt-right.
We analyze two large datasets containing digital traces of these communities, derived from Reddit and YouTube. Overall, our analysis encompasses over 87 million comments across 526 YouTube channels and over 235 million comments and threads made across over 115 subreddits.
These datasets are expanded versions of previously available data~\cite{ribeiro_evolution_2020, ribeiro_auditing_2020}.

Leveraging these comprehensive snapshots, we analyze, within each platform, the similarity between the commenting user bases of the Alt-right and the different communities within the Manosphere.
We compare these results with carefully chosen counterparts: in the case of Reddit, randomly sampled set of users, and in the case of YouTube, users who commented in a predefined set of media channels.
Additionally, we dig deeper and perform a migration study to examine the extent to which users in Manosphere communities gradually engage with Alt-right content, again comparing our results with counterparts.

Overall, we find that for several communities, across platforms, the overlap between users in the Manosphere and the Alt-right is substantial and that it is common for users who once commented exclusively in the Manosphere to eventually engage with Alt-right content. 
These results, however, are heterogeneous within the Manosphere and across the two websites.
For some communities (\eg, MGTOW) a strong link with the Alt-right is observed in both Reddit and YouTube, while for others, this connection stronger in one of the platforms, \eg, users posting in PUA subreddits systematically migrate to Alt-right subreddits but users commenting in PUA YouTube channels do not.

We argue that these findings strengthen the concerns raised by some researchers~\cite{dibranco_male_2020, slpc_male_2018}, and have the potential to guide more fine-grained research into the links between misogynistic and far right movements.
Additionally, we believe that the detailed breakdown of the relationship between different communities within the Manosphere and the Alt-right can empower platforms to craft better policies and take better moderation decisions.\footnote{Code and reproducibility data: \url{https://doi.org/10.5281/zenodo.4420983}.
}

\begin{table*}[t]
    \centering
    \small
    \caption{Brief taxonomy of the communities relevant to this study.
    \vspace{-2mm}}
    \begin{tabular}{p{3.5cm}p{13cm}}
    \textbf{Community} & \textbf{Description}\\ \midrule
    \textbf{Men's Rights Activists \newline (MRA)} & Men's right activists often define themselves as a group focused on men-related social issues and on how institutions consistently discriminate against men~\cite{coston_white_2012}.
    Yet, the movement has been repeatedly called out due to its misogynistic rhetoric~\cite{maddison_private_1999}, and their requests have been labeled ``vengeful'' by some researchers~\cite{mann2008men, dragiewicz2011equality}, demanding, for example, the recognition that men are equally or more victimized by domestic violence than women~\cite{mann2008men}.
    \\ \midrule
    \textbf{Men Going Their Own Way \newline (MGTOW)} & 
    A mostly online community that advocates that men should part ways with women and society, which has been harmed beyond recovery by feminism~\cite{wright_pussy_2020}.
    Ribeiro et al.~\cite{ribeiro_evolution_2020} have shown that, on Reddit, the community was created by users who used to participate in Men's Right Activists subreddits.
    Yet, unlike MRAs, they believe that the system is impossible to change, and the solution is to ``go your own way''~\cite{lin2017antifeminism}.
\\\midrule
    \textbf{Pick Up Artists \newline (PUA)} & Pick up artists are a community centered around the idea of ``the Game,'' strategies that would help men pick up women~\cite{lilly_world_2016}.
    Many of the techniques promoted by members of the community either objectify or harass women.
    For example, they employ a constellation of techniques to deal with ``last-minute resistance to sex,'' encouraging men to bypass signs of lack of consent~\cite{myles_seduction_2019}. 
    \\\midrule
    \textbf{Involuntary Celibates \newline (Incels)} & Incels are a movement, mostly of young men, united by a strong feeling of rejection and rage towards the opposite sex~\cite{tolentino_rage_2018}.
    The community abides by the so-called ``black pill,'' the idea that your looks are the only determinant in a successful romantic life~\cite{incelwiki_blackpill_2021}.
    According to that view, those who do not conform to society's beauty standards are doomed to a life of loneliness and rejection.
    \\\midrule
    \textbf{Alt-right} & A loose segment of the white supremacist movement with a substantial online presence on websites such as 4chan, and in certain corners of YouTube and Reddit~\cite{AltRightAnti-DefamationLeague}. 
    The presence of this community on mainstream platforms has been substantially reduced in recent years due to systematic banning~\cite{hern_reddit_2017,perez_youtube_2020}. \\ \bottomrule
    \end{tabular}
    \label{tab:comms}
\end{table*}

\section{Background \& Related Work}

A growing body of research has studied both the Manosphere and far right movements online. 
We briefly discuss the relevant literature, emphasizing quantitative work done in either Reddit or YouTube (the platforms we analyze in this paper). 
We also provide background information regarding the fringe communities of interest.

We are mainly interested in two sets of communities for the purpose of this work: the Manosphere and the far right.
Both these groups are known to be decentralized and full of contradictions. 
For example, far right groups can be either Christian or Paganistic~\cite{francois_euro-pagan_2007}, and within the Manosphere, some communities perceive a ``crisis in masculinity'' stemming from the feminization of society, while others due to the feminization of men~\cite{lilly_world_2016}.
Also, the nature of the online activity of different communities within the far right and the Manosphere differs. 
For example, far right groups that openly embrace Nazi symbols and ideology are usually not present on websites (\eg, Reddit and YouTube) and prefer anonymous or self-governed platforms such as Telegram channels~\cite{urman_what_2020} or standalone forums~\cite{bowman2009exploring}.
In that context, we limit our scope of research to five communities that are known to prosper (or, in some cases, have prospered) on the platforms analyzed~\cite{ribeiro_evolution_2020, AltRightAlt2019ADL}, which we describe in detail in Table~\ref{tab:comms}.
Four of these communities belong to the Manosphere, and the Alt-right serves as the representative of far right movements.

\xhdr{Far right movements online}
The far right presence online can be traced back to the 1990s, when \emph{Stormfront} emerged, first as a bulletin board, and then as a fully-fledged white supremacist website~\cite{swain_contemporary_2003}.
Yet, it was with the advent of mainstream social networks, that online far right communities, most notably the Alt-right, have gained the spotlight.
The movement had its online presence spread around subreddits (\eg, r/AltRight, r/FrenWorld), YouTube channels (\eg, James Allsup), anonymous imageboards (\eg, 4chan, 8chan), and alternative platforms (\eg, Gab, voat.co). 
The movement was involved in notorious events such as the Christchurch terrorist attack on two mosques in New Zealand~\cite{veilleux2020christchurch}, and the Charlottesville ``Unite the Right'' rally in the United States~\cite{atkinsonCharlottesvilleAltrightTurning2018}.
The Alt-right and fringe platforms associated with it were extensively studied by researchers over the last few years.
Previous research includes large-scale analyses of 4chan~\cite{bernstein_4chan_2011, hine_kek_2017}, Gab~\cite{zannettou_what_2018, mcilroy-young_welcome_2019}, and specific subreddits associated with the Alt-right~\cite{grover2019detecting, flores-saviaga_mobilizing_2018}.
We here make a distinction between Alt-right and Alt-lite~\cite{AltRightAlt2019ADL}.
The term Alt-lite was created to differentiate right-wing activists who deny embracing white supremacist ideology. 
According to Atkinson~\cite{atkinsonCharlottesvilleAltrightTurning2018}, this distinction was exacerbated by the Unite the Right Rally in Charlottesville, which revealed the white supremacist leanings and affiliations to the general public. 

\xhdr{Manosphere} 
The Manosphere and communities within it have been previously studied in Reddit, YouTube, and standalone websites. 
Farell et al.\ have explored the use of misogynistic language in Manosphere-related subreddits~\cite{farrell_exploring_2019} as well as extensively characterized their jargon with computational and socio-linguistic techniques~\cite{farrell_use_2020}.
Ribeiro et al.~\cite{ribeiro_evolution_2020} have mapped the evolution of the Manosphere across Reddit and several standalone websites, showing that newer and more toxic communities such as Men Going Their Own Way and Incels are overshadowing older communities such as Pick Up Artists and Men's Rights Activists.
Studies focusing on specific communities within the Manosphere in a given platform include Papadamou et al.'s investigation of Incel related content on YouTube ~\cite{papadamou_understanding_2021}, and
LaViolette and Hoogan's careful analysis of the platform signals to distinguish the discourse in \url{r/MensRights} and \url{r/MensLib}~\cite{laviolette2019using}, two groups with very different views on masculinity and feminism.
Overall, research on fringe movements and communities (related both to the Manosphere and the far right) stresses that the influence of these communities is not limited to real-world incidents. 
Although ``niche,'' these communities have a disparate influence on our online information ecosystem and are associated with a variety of anti-social behavior online~\cite{flores-saviaga_mobilizing_2018, zannettou_web_2017, dibranco_male_2020}.

\xhdr{Online radicalization}
Much of the research on online radicalization focused on characterizing the activity of Jihadist recruitment and propaganda in mainstream Online Social Networks~\cite{klausen_youtube_2012, klausen_tweeting_2015}.
More recently, radicalization in the context of the Alt-right was also  studied.
As previously mentioned, Lewis suggested that, on YouTube, the proximity between alternative content creators within the platform could create ``radicalization pathways.'' 
These were empirically verified by Ribeiro \etal~\cite{ribeiro_auditing_2020}, who have shown that users that engaged in channels from so-called ``gateway communities'' would consistently migrate to Alt-right content.
The mechanisms governing these radicalization pathways on YouTube have been extensively discussed: while some suggest that YouTube's algorithm would be ``the great radicalizer''~\cite{tufekci_youtube_2018, alfano_technologically_2020}, others have pointed out that radicalization could be driven by novel technological affordances~\cite{munger2020right} and social dynamics on the platform~\cite{lewis_this_2020}.

\begin{table*}[t]
\centering
\caption{Overview of the YouTube data used in this paper. Categories marked with a $*$ were obtained from \cite{ribeiro_auditing_2020}. For brevity's sake, we omit the keyphrases used in the categories from  \cite{ribeiro_auditing_2020}.}

\begin{tabular}{lrrrr|l}
\toprule
Category &    \#Views &   \#Videos & \#Channels & \#Comments & keyphrases\\
\midrule
Alt-lite$^*$ & 4.3B &   69380 & 113 & 48M & --- \\
Alt-right$^*$ & 3.1M &   15375 & 81 & 4.7M & --- \\
Media$^*$ & 14B &  208936 & 66 & 21M & --- \\
Incel & 0.5M &    3526 & 42 & 0.4M & black pill, incels, looksmax \\
MGTOW & 3M &   18749 & 79 & 4.6M & red pill, MGTOW, hypergamy \\
MRA & 1.7M &    5678 & 33 & 1.5M & mra, men's rights, father's rights\\
PUA & 3.1B &   28216 & 124 & 6M & daygame, Indicator of Interest, PUA \\
\midrule
Total & 21.4B &   349860 & 526 & 87M &  \\
\bottomrule
\end{tabular}
\label{tab:yt_descr}
\end{table*}

\section{Data}

Our cross-platform analysis leverages data from both YouTube and Reddit. 
Some of the data employed stems from previous work, while other was collected with methodology mimicking that of previous work. 
Detailed information on our data including the name of the channels and subreddits and data collection logs are provided in an online appendix.\footnote{ \url{https://git.io/JsTm8}}

\subsection{YouTube}

To study YouTube, we expand the dataset obtained from  \cite{ribeiro_auditing_2020}. 
We leverage the same methodology to collect data associated with 4 groups in the Manosphere: Incels, PUAs, MGTOWs, and MRAs. 
The overall data collection process was performed between the 20th and the 28th of May 2019 and is detailed below.

\xhdr{Creating a pool of channels} 
For each of the four communities of interest, we create a pool of candidate channels as follows. 
\textit{1)} We started with a set of seed channels extracted from a variety of sources, blog posts, and wikis.
\textit{2)} Then, we devised a set of key phrases associated with each community (\eg for Incels, the keywords were \textit{black pill}, \textit{incels}, and \textit{looksmax},  we provide a complete list of the keywords used for each community in Table~\ref{tab:yt_descr}).  
For each keyphrase, we use YouTube's search functionality and consider the first 200 results in English. 
We inspected each result and decided whether to add the channel to the pool of candidate channels.
\textit{3)} Finally, we iteratively collected "related" and "featured" channels (a now disabled feature on YouTube), on all the channels obtained in steps (1) and (2); This is done twice.

\xhdr{Annotating the channels} 
Once all candidate channels were collected, we had two annotators label them carefully. 
Annotators had previous experience with the communities at hand, and had to watch, for each channel, at least 5 minutes of content.
The whole annotation period lasted for 3 weeks in May 2019, and the annotator agreement was 91.8\%.
In the end, we obtained a pool of 279 channels, out of which 32 were Incel-related, 80 MGTOW-related, 33 MRA-related, and 124 PUA-related.

\xhdr{Collecting the data}
Lastly, for each channel, we collect the number of subscribers and
views, and for their videos, all the comments.
Notice that we use data along with the YouTube data published by Ribeiro \etal~\cite{ribeiro_auditing_2020}, which was captured in a similar fashion.
We depict statistics for both datasets in Table~\ref{tab:yt_descr}.

\subsection{Reddit}

To study Reddit, we expand the dataset obtained from Ribeiro \etal~\cite{ribeiro_evolution_2020}, where subreddits associated with the Manosphere were collected and curated into the categories of interest. 
We enrich this data with Alt-right-related subreddits using the same methodology.

\xhdr{Creating a pool of (Alt-right) subreddits}
We created a pool of Alt-right subreddits by analyzing \url{r/AgainstHateSubreddit} and \url{r/AntifascistsofReddit}, two communities that oppose hateful groups in the platform.
We manually analyzed all submissions of these two subreddits containing the term "Alt-right," briefly inspected them, and, if appropriate, designated them as candidate channels.
Additionally, we also collect two sets of counterparts. 
First, we collect all comments and submissions belonging to 17 popular Gaming subreddits: gaming online communities are known to be male-dominated and provide a good point of comparison~\cite{massanari_gamergate_2017}. 
These were randomly sampled from the list of top 100 gaming subreddits available in the \url{r/gaming} community\footnote{\url{https://www.reddit.com/r/gaming/wiki/list-sorted-by-subscribers}}.
Second, we collect a random sample of 0.5\% of all Reddit posts available through Pushshift. 
As we discuss later, these two counterparts will each be used in a different analysis. 

\xhdr{Annotating the subreddits}
After creating the pool of 65 subreddits, we spent a week in January 2021 carefully inspecting each one of them for at least 3 minutes and annotating them as either belonging to the Alt-right or not. 
When subreddits were banned (which was often the case), we resorted to the Internet Archive to browse historical snapshots. 
Note that all of these  subreddits were available in the Pushshift dataset. 

\xhdr{Collecting the data} 
We ended up with a pool of 59 subreddits, from which we collected all available data using Pushshift~\cite{baumgartner_pushshift_2020}. 
Notice that we use this data along with the Reddit data published with \cite{ribeiro_evolution_2020}. 
We depict statistics for both datasets in Table~\ref{tab:rd_descr}.

\begin{table}[t]
\centering
\caption{Overview of the Reddit data used in this paper. Categories marked with a $*$ were obtained from \cite{ribeiro_evolution_2020}.}
\begin{tabular}{lrrrr}
\toprule
Category &  \#Subred. &  \#Users &   \#Threads & \#Posts \\ \midrule
Incel$^*$ &      18 &    197,194 &    349,711 &   5,075,962 \\
MGTOW$^*$ &       4 &     85,293 &    183,757 &   3,243,062 \\
MRA$^*$   &      10 &    214,011 &    219,809 &   3,989,282 \\
PUA$^*$   &       7 &    177,093 &    203,749 &   1,664,341 \\
Alt-right &      59 &  1,136,500 &  4,631,572 &  55,345,906 \\
Random    & - &  7,623,391 &  2,896,690 &  29,921,245 \\
Gaming    &      17 &  3,144,631 &  6,560,103 & 121,399,219 \\
\midrule
Total     & 115 & 12,578,113 & 15,045,391 & 221,173,017\\
\bottomrule
\end{tabular}
\label{tab:rd_descr}
\end{table}

\begin{figure*}[t]
    \centering
    \includegraphics[width=\linewidth]{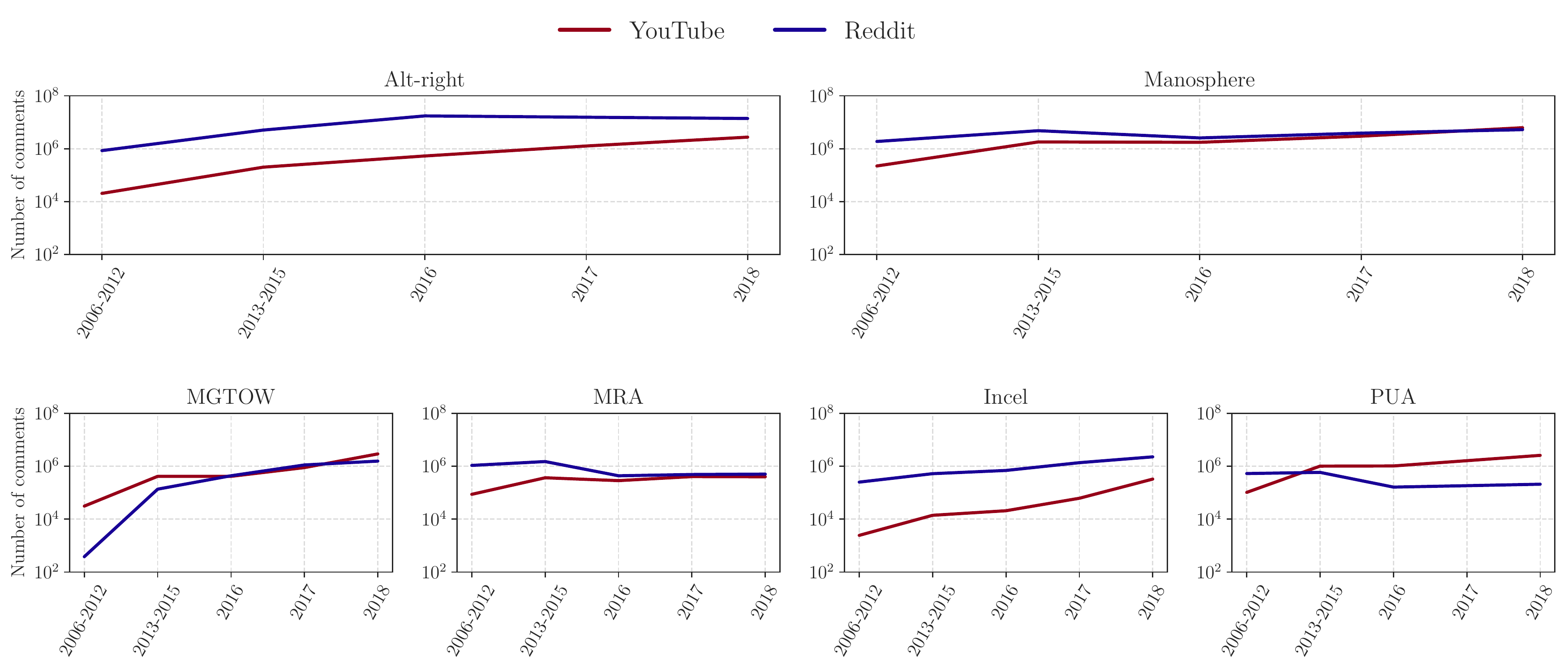}
    \caption{\textbf{Overview of commenting activity:} We depict, for all communities of interest, the number of comments per community on YouTube and Reddit for the indicated periods of time. 
    Notice that we present Manosphere communities both in aggregate (in the upper row) and separately (in the bottom row). 
    For the Alt-lite and the Intellectual Dark Web, we consider only YouTube data.}
    \label{fig:ov}
\end{figure*}

\subsection{A note on counterparts and other gateway communities}

Our analyses employ counterparts: other groups of users which we use for comparison.
For YouTube, we use the channels classified as ``media'' by Ribeiro et al.~\cite{ribeiro_auditing_2020}, and for Reddit, we use a set of gaming subreddits and a set of random comments.
We avoid referring to these comparison groups as ``controls,'' due to the term's causal nature, which does not hold in the observational setting we study.
Nevertheless, as indicated by Rosenbaum~\cite{rosenbaum2017observation}, ``if counterparts are governed by the same laws or forces as treated and control groups, we may study the operation of those forces in the absence of treatment by studying the counterparts.''

In that sense, we argue that counterparts provide a sanity check on the effect sizes observed across different communities and different platforms. 
If we observe substantially more overlap or migration between the communities of the Manosphere and the Alt-right than between counterparts and the Alt-right, we argue that this constitutes good evidence of close ties between the Manosphere and the Alt-right.
Here, it also helps that we are studying two independent platforms:  we can be more confident of findings that are ``reproducible'' both on Reddit and on YouTube.

Additionally, in our analyses leveraging YouTube data, we also compare the ties between the Manosphere and the Alt-right with the ties between the Alt-lite and the Alt-right. 
Migration between the Alt-lite and the Alt-right has been previously studied by Ribeiro et al.~\cite{ribeiro_auditing_2020} and this comparison provides yet another way to assess the effect size of our results.
If counterparts allow us to compare the Manosphere with communities where we would expect weak ties with the Alt-right, the Alt-lite allows comparing the Manosphere to a well-known gateway to the far right.

\subsection{Overview of commenting activity in the communities of interest}

\Figref{fig:ov} depicts the number of comments in each of the communities in both YouTube and Reddit.
Across the two platforms, the number of comments is very similar for some of the communities within the Manosphere (\eg, MGTOW, MRA), while for others, they differ by orders of magnitude (\eg, PUA, Incels).
Additionally, we find the growth of these different communities varies substantially: while some communities grew substantially during the study period (\eg MGTOW), others remained mostly stagnated (\eg MRA).

\section{User Base Similarity}
\label{ch:sim}

\begin{figure*}[t]
    \centering
    \includegraphics[width=\linewidth]{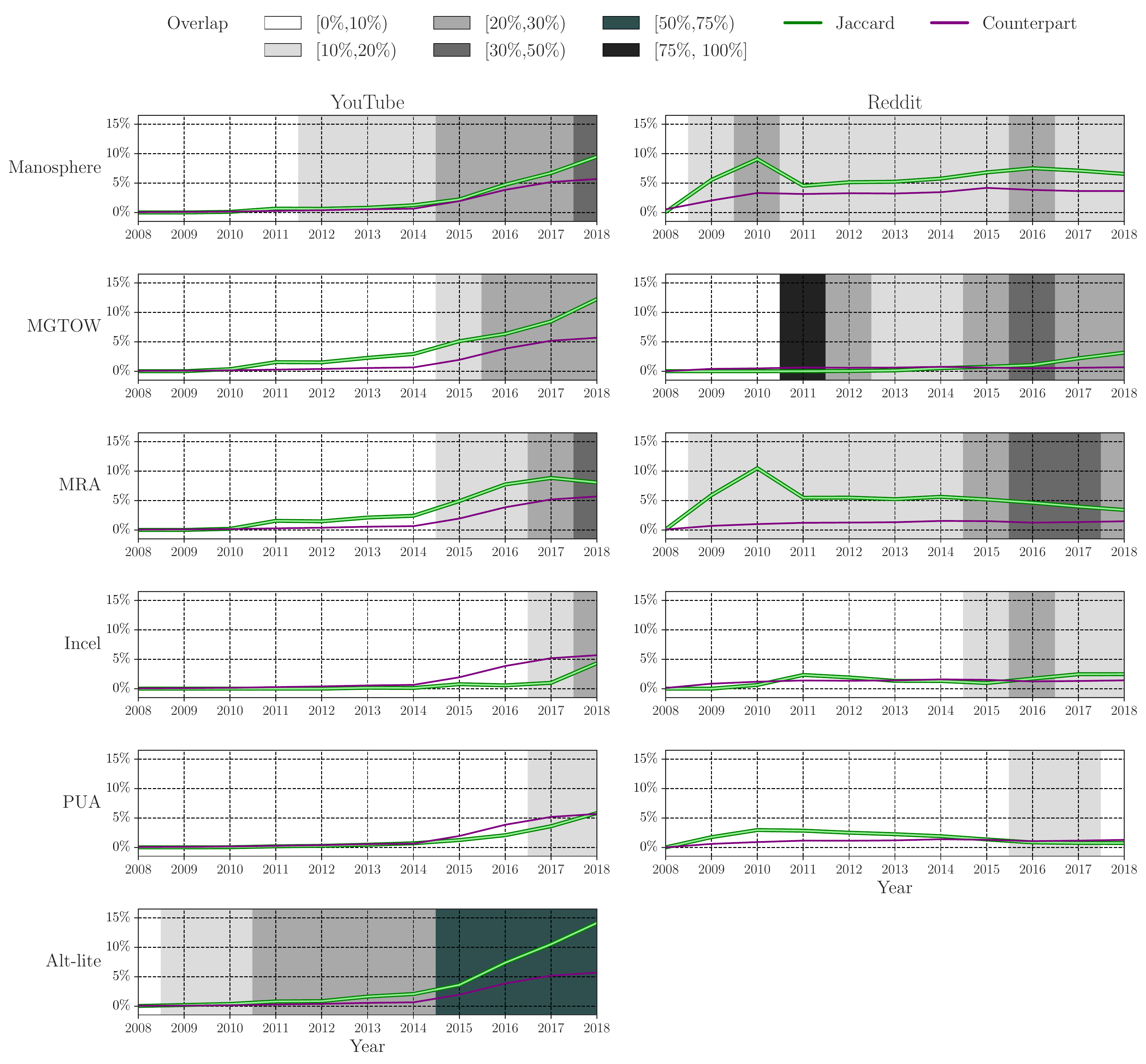}
    \caption{\textbf{Similarity between the user base of the Manosphere and of the Alt-right:}
    We show the evolution of the Jaccard similarity (the green line) and overlap coefficient (indicated by the background color) between the studied fringe communities ---named on the left--- and the Alt-right for both YouTube (on the left) and Reddit (on the right).
    We additionally show the Jaccard similarity of the counterparts (the purple line).
    Notice that here too, we analyze the Manosphere in aggregate (on the first row) and split into four communities (in rows 2-5).
    We also depict the similarity between the Alt-Lite and the Alt-right.
    This is done only for YouTube as we do have not collected Alt-lite subreddits.
    }

    \label{fig:sim}
\end{figure*}

We study the user base similarity between different communities using two metrics: the Jaccard similarity  $\frac{|A \cap B|}{|A \cup B|}$ and the overlap coefficient $\frac{|A \cap B|}{\min(|A|,|B|)}$, where A and B are the set of commenting users in two communities.
The overlap coefficient yields fairer results when comparing communities of different sizes since a community can be entirely contained within another one and yield a low Jaccard similarity score.
\Figref{fig:sim} compares the user base of the Alt-right and the communities in the Manosphere across different years.
We provide two additional points of comparison.
First, on YouTube, we depict the similarity between the Alt-right and the Alt-lite, a known gateway community~\cite{ribeiro_auditing_2020}.
Second, we report the similarity between each community and the chosen counterparts.
For YouTube, we use the media channels collected by Ribeiro \etal~\cite{ribeiro_auditing_2020}.
For Reddit, we use the random sample of comments collected from Pushshift.
Since the number of distinct users is substantially bigger in the random sample, we subsample it to match the size of the community we are comparing it with.

\xhdr{The Manosphere and the Alt-right}
The first row of \Figref{fig:sim} shows the similarity between the Manosphere and the Alt-right in both YouTube (left) and Reddit (right).
We observe that the Jaccard similarity between the Alt-right and the Manosphere is higher than the one between the Alt-right and the counterpart on YouTube.
The gap between the two, as well as the overlap coefficient between the Manosphere and the Alt-right, is also growing in recent years.
For example, in 2016, the Jaccard similarity was 4.7\% and the overlap 22.5\%, while in 2018, these figures have grown to 9.5\% and 31.7\%, respectively.
However, we find these numbers to be less expressive than for the Alt-lite, a previously known gateway community on YouTube (shown in the sixth row). 
For example, in 2018 the Alt-lite had a Jaccard Similarity score of 14.1\% and overlap coefficients of 70.5\%.
On Reddit, we also observe that the Manosphere is consistently more similar to the Alt-right than the randomly sampled counterpart. 
Here, the Jaccard similarity has local maxima in 2010 (9\% \vs 3.1\% counterpart) and 2016 (7.5\% \vs 3.8\%). 
In both these years, the overlap coefficient was bigger than 20\%.

\xhdr{Inside the Manosphere}
We extend our analysis by analyzing the Manosphere communities separately, rather than in aggregate (as shown in rows 2 to 5).
Here, we find that communities behave heterogeneously.
On the one hand, in recent years, MGTOWs and MRAs are more similar to the Alt-right than the counterparts on both YouTube and Reddit.
For example, in 2018 MGTOWs had Jaccard similarity scores of 3.2\% on Reddit (\vs 0.05\% counterpart) and 12.2\% on YouTube (\vs 5.6\% counterpart).
On the other hand, when comparing PUAs and Incels and the counterparts in both Reddit and YouTube, we find either small differences or that counterparts have higher similarity scores than the Manosphere communities.
Importantly, notice that this is not merely a consequence of the size of these communities: on YouTube, for example, PUAs were the largest of all communities (as measured by the number of comments), while Incels were the smallest (see \Figref{fig:ov}).
Lastly, we also notice that in the founding years of the MGTOW community (when it still was quite small, with only a couple of hundreds of posts, as shown in \Figref{fig:ov}), its overlap with the Alt-right was particularly strong on Reddit. 
In 2012, for example, 17 of the 72 active MGTOW members were also active in Alt-right subreddits (23.6\% overlap).
This suggests that the founding members of the MGTOW community on Reddit were also a part of the Alt-right community.

\section{User Migration}
\label{ch:mig}

In the previous section, we showed that,  in recent years, the commenting user bases of the Manosphere ---in particular of the MGTOW and the MRA communities--- are becoming increasingly similar to those of the Alt-right. 
This suggests that there is a growing percentage of users consuming both Manosphere \textit{and} Alt-right content. 
Yet, this observation does not conclusively indicate if the Manosphere is a ``gateway'' to the Alt-right: users could join both communities at roughly the same time or migrate from the Alt-right to the Manosphere.
To gain more insight into this, we conduct a more fine-grained study.
We trace the comment history of YouTube and Reddit users by tagging them in a community in a given period and follow their actions into the future.
To observe migration patterns, we analyze the proportion of members active in a given community at a given time that engaged with Alt-right content in the following years.

\begin{figure*}[t]
    \centering
    \includegraphics[width=\linewidth]{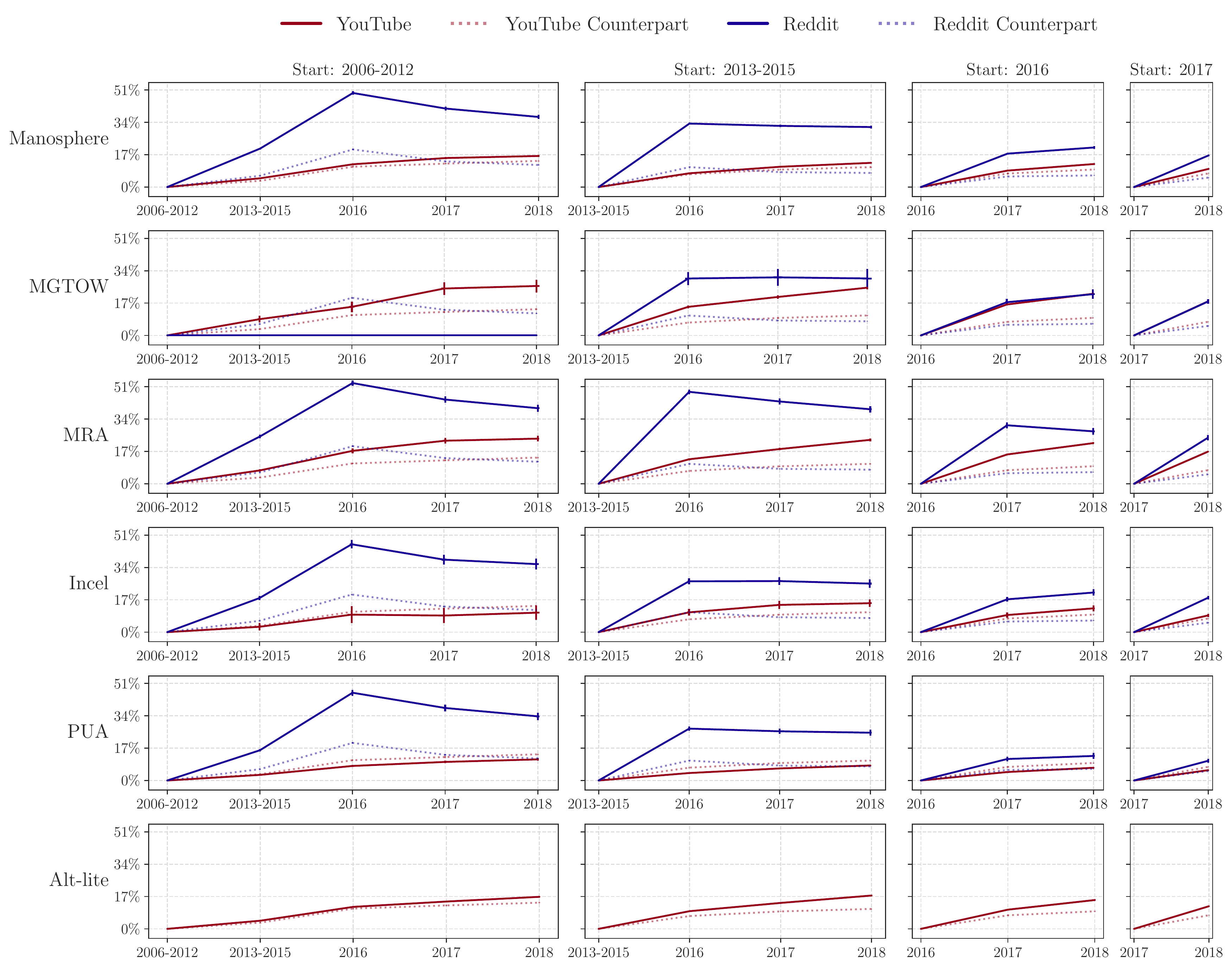}
    \caption{\textbf{Migration of users towards Alt-right content:}
    We report the results of a migration analysis performed in all communities of interest. 
    We track how users who commented exclusively in a given community (one per row) in a given starting period (one per column) eventually engage with Alt-right content in years to come on both Reddit (in blue) and YouTube (in red).
    Here we use as media channels as YouTube counterparts and gaming subreddits for Reddit (shown as blue and red dotted lines). 
    95\% confidence intervals are shown by (tiny) error bars.
    The plot tracking MGTOW in the period 2006--2012 on Reddit is devoid of any data points because almost no user is tracked for this starting period.}
    \label{fig:mig}
\end{figure*}

The results for this analysis are shown in \Figref{fig:mig}. 
Each column corresponds to a different starting point (2006 to 2012, 2013 to 2015, 2016, or 2017), and each row corresponds to a different community.
In each plot, we track all users who, in a given starting point, commented in the community indicated by the row and not in the Alt-right.
We then show, for subsequent periods, what percentage of those users kept posting on the platform \emph{and} engaged with either Alt-right channels (in red) or subreddits (in blue).
Importantly, we compare our results with counterparts.
For YouTube, we again use the media channels obtained from Ribeiro \etal~\cite{ribeiro_auditing_2020}. 
For Reddit, we use the set of 17 gaming subreddits. 
We avoid using the random sample of posts here: since sampling was done at the post level, the data does not contain complete longitudinal data for individual users, which makes the data inappropriate for this analysis.

We further clarify this methodology with an example. 
Consider the plot corresponding to the MGTOW community with a starting point in 2016 (second row, third column).
First, we select all users who commented in the MGTOW community in 2016 (and not in the Alt-right).
Then, for 2017 and 2018, we calculate the fraction of users that went on to comment in Alt-right channels or subreddits. 
On YouTube (solid red line) we find that from the set of users who commented in the MGTOW community in 2016, 16.2\% and 21.9\% of those who remained active went on to engage with Alt-right subreddits in 2017, and 2018 respectively.
These numbers are significantly larger than the migration observed in the counterpart (dashed red line): for users who in 2016 commented exclusively on media channels and remained active in subsequent years, less than 10\% went on to comment in Alt-right channels.

\xhdr{Reddit}
Analyzing the migration from the Manosphere towards the Alt-right in Reddit (shown in the first row of \Figref{fig:mig}), we 
find that, across the different starting periods, Reddit users of the analyzed Manosphere communities tend to drift towards Alt-right subreddits significantly more than the counterpart.
This behavior is valid for all studied communities and all starting periods but is the strongest for MRAs.
For example, for users who exclusively commented in MRA subreddits in 2017, around 24\% of those who remained active in 2018 went on to comment in Alt-right subreddits (\vs\ 5\% for the counterpart).
Considering that the user bases of PUAs and Incels are, in Reddit, less similar to the user base of the Alt-right than MRAs and MGTOWs, it is noteworthy that members of all four manosphere communities migrate to Alt-right subreddit at similar rates.
This suggests that having similar user bases and being a gateway to other communities are distinct phenomena.
Previous work has suggested that there is widespread migration also within the Manosphere~\cite{ribeiro_evolution_2020}, and it could be that users ``indirectly'' migrate to the Alt-right. 
For example, PUA members might join the MGTOW community first and then join the Alt-right later.

\xhdr{YouTube}
On YouTube, we find mixed results: although communities within the Manosphere migrate to the Alt-right more than the counterparts, the differences between the two groups are much smaller (\Figref{fig:mig}, first row). 
For example, for users which we track from 2006--2012, we observe around 15.7\% of engagement from uses who initially commented on the Manosphere (\vs 13.4\% for the counterpart).
Although the result is statistically significant (as shown in the CIs), the gap is much smaller than what we observed for Reddit, where the migration to the Alt-right from the Manosphere is roughly twice the migration to the Alt-right from the counterpart. 
These differences between the observed values and the counterparts cannot be explained solely by the nature of the websites---forum rather than video platform---because our chosen counterparts exhibit very similar patterns on both platforms.
When analyzing communities within the Manosphere separately, we find that members of the Incel and the PUA communities migrate less to the Alt-right than members of the MRA and MGTOW communities.
For example, starting from 2013--2015, we find that 25.1\% of users who initially commented on the MGTOW community migrated to the Alt-right, whereas only 15.2\% of users who initially commented on the Incel community did so.
This finding is more aligned with what we observe in \Figref{fig:sim} and suggests that, on YouTube, the link between the PUA and Incel communities and the Alt-right is weaker.

\xhdr{Alt-lite}
We can also use the migration between the Alt-lite and the Alt-right as another point of comparison (again, only for YouTube). 
Considering communities in the Manosphere in aggregate, we find that Manosphere users migrate to the Alt-right less often than Alt-lite users.
For example, for users starting in 2016 in the Manosphere, 12.7\% of them eventually migrate to the Alt-right (\vs~15.1\% for the Alt-lite).
However, considering Manosphere communities separately, we find that MRAs and MGTOWs migrate to the Alt-right significantly more often than members of the Alt-lite.
For instance, out of the users starting in 2016 in the MGTOW community, 25.1\% would eventually migrate to the Alt-right (\vs 15.1\% for the Alt-lite).

\section{Discussion}
\label{ch:discussion}

Our analysis suggests that, in both YouTube and Reddit, there is significant overlap in the user bases of the Manosphere and of the Alt-right (\Secref{ch:sim}) and that users in the Manosphere systematically go on to consume Alt-right content (\Secref{ch:mig}).
This provide quantitative evidence to the link between the Manosphere and the Alt-right, which has been hypothesized by researchers and NGOs in recent years~\cite{slpc_male_2018, dibranco_male_2020}.
In what follows, we discuss nuances, implications, and limitations of the work at hand.

When interpreting the results presented here, it is important to keep in mind that these are ever-changing communities and that tracing boundaries is often tricky.
Here, we study the links between the communities in Manosphere and the Alt-right. 
However, it could be argued that the term Alt-right has since become obsolete with the rise of new groups such as the Boogaloo and Q-Anon movements.
It would be interesting for future work to continue to characterize migratory flows between these ever-evolving fringe groups.

Another limitation of this work is that we focus on studying migratory patterns without trying to explain their causes. 
In that context, future work could try to explain \textit{why} users migrate to the Alt-right. 
This could be done a qualitative level, by taking a closer look at the data, but also through large-scale analyses: for example, one could try to predict whether a user will do so by the digital traces they produce, such as the language they use.\footnote{This direction has been explored on Reddit, but not in the context of community migrations \cite{grover2019detecting}.}

Within the Manosphere, we observe that different communities have different relationships with the Alt-right. 
This finding not only resonates with previous work differentiating these communities by their linguistic traces~\cite{farrell_use_2020}, but also suggests that not all Manosphere communities are equally problematic. 
We find that, on both Reddit and YouTube, Men Going Their Own Way (MGTOW) and Men's Rights Activists (MRA) were consistently linked with the Alt-right, both in terms of their user-base similarity and in terms of the migratory patterns observed.
This is not the case for the other two Manosphere communities studied, Incels and Pick Up Artists (PUA).
This finding is in stark contrast with the disparate focus on the alleged pipeline from Incels to the Alt-right ~\cite{szalai_undercover_2020, papadamou_understanding_2021}. 
While the ties between Incel ideology and male supremacist terrorist attacks~\cite{dewey_inside_2014} should continue to be explored, our research suggests that stakeholders should also investigate influence of MGTOWs and MRAs on our online information ecosystem.

A second implication of the work at hand has to do with the role of recommender systems in online radicalization, a topic that has been extensively discussed in previous research~\cite{munger2020right, ribeiro_auditing_2020}.
Here, we find that there was migration from Manosphere communities both on YouTube, a platform largely driven by automatic recommendations, and on Reddit, a platform governed by more transparent mechanisms (\eg, upvote, karma, \etc). 
Also, we find that migration across all of the communities in the Manosphere to the Alt-right is substantially more prevalent on Reddit than on YouTube.
We argue that this constitutes evidence towards the hypothesis that online radicalization can happen independently of recommender systems (as suggested for example, by Munger and Phillips~\cite{munger2020right}).

\bibliographystyle{ACM-Reference-Format}
\bibliography{bib}

\end{document}
\endinput